\documentclass[%
reprint,
 amsmath,amssymb,
 aps,
pra,
longbibliography,
]{revtex4-2}
\usepackage{graphicx}% Include figure files
\usepackage{dcolumn}% Align table columns on decimal point
\usepackage{float}
\usepackage{bm}% bold math
\usepackage{pst-node}
\usepackage{tikz-cd} 

\begin{document}

\preprint{APS/123-QED}

\title{Energy Spectrum of a constrained Quantum Particle \\ and the Willmore Energy of the constraining Surface}

\author{Vicent Gimeno i Garcia}
\email{gimenov@uji.es}
 \affiliation{Departament de Matem\`{a}tiques- IMAC,
Universitat Jaume I, Castell\'{o}, Spain.}%Lines break automatically or can be forced with \\

\author{Steen Markvorsen}%
 \email{stema@dtu.dk}
\affiliation{%
 DTU Compute, Technical University of Denmark, Lyngby, Denmark
}%

\date{\today}% It is always \today, today,

\begin{abstract}
Geometric and topological bounds are obtained for the first energy level gap of a particle constrained to move on a compact surface in 3-space. Moreover, geometric properties are found which allows for stationary and uniformly distributed wave functions to exist on the surface. 

PACS number(s): 03.65.Ge, 68.65.2k

\end{abstract}

\maketitle

%\tableofcontents

\section{\label{sec:level1}Introduction }
During the first and second decades of this century the study of surface physics has attracted attention from both theoretical and experimental perspectives. Nanostructures, such as fullerenes, carbon-based low-dimensional shapes, and nanotubes have been researched due to their potential technological applications (see references \cite{Zheng20101647,Li2020, Ortix2010,Sadykov2020369, Cheng2021,Gravesen2005105,Gravesen2010, Gravesen2005,Gravesen20201}).

From a theoretical perspective, as developed e.g. in \cite{DeWitt1957377}, the dynamics and energy levels of a quantum particle on a surface can be described as a particle in the curved space defined by the surface, and can thus be viewed from a purely intrinsic perspective. However, when considering a particle that is confined to an immersed surface in a locally Euclidean 3-space, an extrinsic perspective can also be adopted.

In fact, according to \cite{Jensen71} and \cite{DaCosta}, an extrinsically motivated effective potential is present in the Schrödinger equation for a surface in 3-space. That potential involves both the Gaussian curvature (i.e. an intrinsic property of the surface) and the mean curvature (an extrinsic property of the surface). This idea of inducing a combination of extrinsic and intrinsic contributions into the governing equations for the quantum phenomena on a surface can be further extended, for example by adding electromagnetic fields or in the setting of Quantum Field Theory (as discussed in the following references \cite{Jensen2011448, Ortix2011,Matsutani1993686,Burgess19931861, Jensen2009, Homma19902049, Ferrari2008} for instance). 

 Da Costa in \cite{DaCosta}, and first Jensen-Koppe in \cite{Jensen71}, considered  the classical Schrödinger equation for a quantum scalar particle confined to a surface $\Sigma$ in $\mathbb{R}^3$ and obtained an equation for the particle which contains an effective potential expressed in terms of the mean curvature and the Gaussian curvature of the surface. More precisely, the classical Schrödinger equation was first considered in  a tubular neighbourhood of the the surface in $\mathbb{R}^3$:
$$
\Sigma_\epsilon =\{x\in \mathbb{R}^3\,  :\, {\rm dist}(x,\Sigma)\leq \epsilon\}.
$$ 
Assuming that the surface $\Sigma$ is compact, regular and embedded without boundary the Schrödinger equation is -- for sufficiently small $\epsilon$ -- well defined in the tubular 3-dimensional domain $\Sigma_\epsilon$, when this domain is equipped with Dirichlet boundary conditions. After shrinking $\epsilon\to 0$, da Costa then obtained the following Schrödinger equation for the particle confined in the surface $\Sigma$,
\begin{equation}\label{extrinsic}
i\hbar\frac{\partial}{\partial t}\Psi=\frac{\hbar^2}{2m}\left[-\Delta -\frac{1}{4}\left(k_1-k_2\right)^2\right]\Psi,
\end{equation}
where $k_1$ and $k_2$ denote the principal curvatures of $\Sigma$. An effective potential $V_{\rm eff}$ appears in this way  with the following alternative geometric expression
$$
V_{\rm eff}=-\frac{\hbar^2}{2m}\left(H^2-K\right)=-\frac{\hbar^2}{8m}S^2,
$$
where $H=(k_1+k_2)/2$ is the mean curvature,  $K=k_1k_2$ is the Gaussian curvature and $S=k_1-k_2$ is the skew curvature. The skew curvature measures how far the surface is from being umbilical. See the introduction in \cite{Lopez2020} for a well motivated description of the skew curvature and \cite{daSilva2021317} for the extension of the skew curvature for surfaces in the Lorentz-Minkowski space and the interpretation of the skew curvature in terms of the standard deviation of the normal curvature seen as a random variable.

Although equation \eqref{extrinsic} is commonly seen in the context of embedded, compact regular surfaces, where it has direct and significant physical implications, it is also an interesting and mathematically well-defined problem to examine the development of wave functions in terms of equation \eqref{extrinsic} for compact regular surfaces that are merely immersed -- not necessarily embedded. Thus, in this paper, we will consider general immersed surfaces with the Schrödinger equation \eqref{extrinsic}.

To find the energy spectrum of the particle, the eigenvalue problem for the corresponding Hamiltonian has to be solved, {i.e.} we have to analyze the family of  stationary wave functions $\{\Psi_k\}_{k\in \mathbb{N}}$ and the energy spectrum $\{E_k\}_{k\in \mathbb{N}}$ such that
\begin{equation}\label{spectrum}
\frac{\hbar^2}{2m}\left[-\Delta -\frac{1}{4}S^2\right]\Psi_k=E_k\Psi_k.
\end{equation}

In this report we focus on the problem of obtaining upper bounds for the gap in the energy spectrum (see inequalities \eqref{result1} and \eqref{result2}) using topological and conformal invariants for the surfaces in question. The paper is organized as follows: in Sect. II, the
Willmore energy and the Euler characteristic will be explained. These fundamental concepts will be the conformal and topological invariants that we will use to obtain the upper bounds for the gap in the spectrum of the energy of a particle confined in a given surface. In Sect. III, we prove inequality \eqref{result1}, which gives an upper bound for the difference of the two first bounded energy levels $E_0$ and $E_1$. Sect. IV is devoted for the study of $E_k-E_0$ for $k>1$ and we obtain the qualitative inequality \eqref{result2}. In Sect. V, we give a simple construction and description of the family of surfaces of revolution with constant skew curvature. In Sect. VI, we show how our techniques and results can be extended to  non-compact surfaces when they satisfy certain symmetry conditions. Sect. VII is devoted to concluding remarks. 

\section{Willmore energy and Euler Characteristic}
The Willmore energy of an immersed surface is a conformal invariant (see \cite{Toda20171}) and is defined as the integral of the square of the mean curvature over the surface,
$$
\mathcal{W}(\Sigma)=\int_{\Sigma}H^2dA.
$$
It is known to be always greater than or equal to $4\pi$, with equality only for the round spheres. Moreover, the Gauss-Bonnet Theorem relates the integral of the Gaussian curvature over the surface to the Euler characteristic, which is a topological invariant in the following way:
$$
\int_\Sigma K\, dA=2\pi \chi(\Sigma).
$$
The Euler characteristic $\chi(\Sigma)$ of an oriented and compact surface without boundary is either equal to $2$ (for the sphere and for any surface homeomorphic to the sphere) or equal to $2-2g$, where $g$ is the genus of the surface.
\section{First Gap in the spectrum of energy and Willmore energy} In this section we provide an upper bound for the difference of the first and second energies of a quantum particle in a two-dimensional surface with mean curvature $H$, Gauss curvature $K$,  and skew curvature $S$. The Hamiltonian $\frac{\hbar^2}{2m}\left(-\Delta-V\right)$, with $V = H^2-K$, describes the system and its spectrum of energies will be denoted by $\{E_i=\frac{\hbar^2}{2m}\lambda_i\}_{i\in\mathbb{N}}$.
Equation (\ref{spectrum}) can be expressed as 
$$
\left[-\Delta-V\right]\Psi=\frac{2mE}{\hbar^2}\Psi,\quad V=H^2-K.
$$
Let $\{\lambda_i\}_{i\in\mathbb{N}}$ be the spectrum of the operator $-\Delta-V$, and  hence,  $E_i=\frac{\hbar^2}{2m}\lambda_i$. By Theorem 2.1 of \cite{ElSoufi2000},
$$
\begin{aligned}
\lambda_1\leq &\frac{1}{{\rm A}(\Sigma)}\int_\Sigma \left(2H^2-H^2+K\right)d{\rm A}\\
=&\frac{2\mathcal{W}(\Sigma)}{{\rm A}(\Sigma)}-\frac{1}{4}\overline{S^2} ,
\end{aligned}
$$  
where $\lambda_1$ is the second eigenvalue of the operator $-\Delta-V$, ${\rm A}(\Sigma)$ is the area of $\Sigma$, and $\overline{S^2}:=\frac{\int_\Sigma S^2d{\rm A}}{{\rm A}(\Sigma)}$ is the mean of the squared skew curvature of $\Sigma$. The first eigenvalue $\lambda_0$ of the operator $-\Delta-V$ can also be estimated from below. Let $u$ be an eigenfunction associated with $\lambda_0$. Then using the Rayleigh quotient we get
$$
\begin{aligned}
\lambda_0=&\frac{-\int_\Sigma u\Delta u dA-\int_\Sigma Vu^2dA}{\int_\Sigma u^2dA}\\=&\frac{\int_\Sigma\Vert \nabla u\Vert^2 dA}{\int_\Sigma u^2dA}-\frac{\int_\Sigma Vu^2dA}{\int_\Sigma u^2dA}\\
&\geq -\max_{\Sigma}V=-\frac{1}{4}\max_{\Sigma}S^2.
\end{aligned}
$$
Finally, since $E_i=\frac{\hbar^2}{2m}\lambda_i$, we can state the following upper bound for the energy difference:
%\begin{widetext}
\begin{equation}\label{result1}
\begin{aligned}
E_1-E_0\leq  &  \frac{\hbar^2}{2m}\left(\frac{2\mathcal{W}(\Sigma)}{{\rm A}(\Sigma)}+\frac{1}{4}\left(\max_\Sigma S^2-\overline{S^2}\right) \right) .    
\end{aligned}
\end{equation}
%\end{widetext}
To achieve this upper bound, it is essential that each inequality employed in its derivation is transformed into an equality. Specifically, a necessary condition for achieving this upper bound is that the surface must be a surface with constant skew curvature.  Additionally, the upper bound  given by equation \eqref{result1} is sharp (the equality is achieved for at least one surface). This is because for a sphere with a radius of $R$ in $\mathbb{R}^3$, it is known that $\mathcal{W}=4\pi$, ${\rm A}(S_R)=4\pi R^2$, and $S=0$. Therefore, our upper bound given in inequality \eqref{result1} can be expressed as follows:
$$
0\leq E_1-E_0\leq \frac{\hbar^2}{2m}\cdot \frac{2}{R^2}
$$
However, it should be noted that in this particular case, since $H^2-K=0$, the spectrum is equivalent to the Laplacian spectrum. Hence, according to \cite{Chavel}, we know that:
$
E_1-E_0=\frac{\hbar^2}{2m} \frac{2}{R^2}.
$
As far as we know it is still unknown if inequality \eqref{result1} becomes an equality for other surfaces aside from the round sphere. Nevertheless, as we mentioned before, a prerequisite for achieving equality is that the surface must have  constant skew curvature.

 We must notice that $\max_\Sigma S^2-\overline{S^2}\geq 0$ with equality only for  surfaces of constant skew curvature and moreover by using the Gauss-Bonnet Theorem we have
$$
\frac{1}{4}\overline{S^2}=\frac{\int_\Sigma H^2-K}{{\rm A}(\Sigma)}=\frac{\mathcal{W}(\Sigma)-2\pi \chi(\Sigma)}{{\rm A}(\Sigma)}.
$$
Therefore, for orientable surfaces:
$$
\frac{1}{4}\overline{S^2}=\frac{\mathcal{W}(\Sigma)-4\pi(1-g)}{{\rm A}(\Sigma)}\cdot
$$
\section{Further gaps in the spectrum}
To provide an upper bound for the difference between consecutive eigenvalues $E_{k-1}$ and $E_0$ we will analyze the number of non-positive eigenvalues $\mathcal{N}(V_\alpha)$ of the Schrödinger operator $-\Delta-V_\alpha$, with $V_\alpha=H^2-K+\alpha$, with $\alpha$ being a real constant.

The number of non-positive eigenvalues of the Schrödinger operator $-\Delta-V_\alpha$ have been bounded from below by 
$$
\mathcal{N}(V_\alpha)\geq c(g)\int_{\Sigma}V_\alpha dA
$$
in \cite{GriNar2016}, where $c(g)$ depends only on the genus of $\Sigma$. By choosing $\alpha$ such that the lower bound is $k$, \emph{i.e.},
\begin{equation}\label{keq}
 c(g)\int_{\Sigma}V_\alpha dA=k ,  
\end{equation}
the first $k$-th  eigenvalues $\{\lambda_i(V_\alpha)\}_{i=0}^{k-1}$ associated with $-\Delta-V_\alpha$ are non-positive 
$$
\lambda_0(V_\alpha)\leq \lambda_1(V_\alpha)\leq \cdots \leq \lambda_{k-1}(V_\alpha)\leq 0.
$$
Therefore an upper bound of $\lambda_{k-1}-\lambda_0$ can now be derived and expressed in terms of $\alpha$ and $\overline{S^2}$  because  
\begin{equation}\label{eqkvina}
    0\leq \lambda_{k-1}(V_\alpha)-\lambda_0(V_\alpha)\leq -\lambda_0(V_\alpha)=-\lambda_0+\alpha. 
\end{equation}
But applying equation (\ref{keq})
$$
\begin{aligned}
k=c(g)\int_{\Sigma}V_\alpha dA=&c(g)\left(\int_{\Sigma}VdA+\alpha {\rm A}(\Sigma)\right)\\
=&c(g) \left(\frac{1}{4}\int_\Sigma S^2 d{\rm A}+ \alpha{\rm A}(\Sigma)\right).
\end{aligned}
$$
 Therefore, the parameter $\alpha$ can be obtained as
\begin{equation}\label{alpha}
\alpha=\frac{k}{{\rm A}(\Sigma) c(g)}-\frac{1}{4}\overline{S^2}.
\end{equation}
By using this expression for the value of $\alpha$, the upper bound of inequality (\ref{eqkvina}) can be written as
$$
\begin{aligned}
\lambda_{k-1}-\lambda_0\leq &-\lambda_0+\frac{k}{{\rm A}(\Sigma) c(g)}-\frac{1}{4}\overline{S^2}.
\end{aligned}
$$
But we can estimate  $\lambda_0$ from below as before and get
$$
\begin{aligned}
\lambda_0
&\geq -\frac{1}{4}\max_{\Sigma}S^2.
\end{aligned}
$$
Then,
%\begin{widetext}
\begin{equation}\label{result2}
 E_{k-1}-E_0\leq \frac{\hbar^2}{2m}\left(\frac{k}{{\rm A}(\Sigma) c(g)}+\frac{1}{4}\left(\max_{\Sigma}S^2-\overline{S^2}\right)\right). 
\end{equation}
Although the theoretical constant $c(g)$ has not yet been determined, and its exact value is unknown, it is not possible to determine the equality case. However, we do know that when equality is achieved, as in the case of inequality \eqref{result1}, the surface must be a surface of constant skew curvature.

 Inequality \eqref{result2} leads to the Weyl's type asymptotic formula 
 $$\liminf_{k\to \infty}\frac{E_k-E_0}{k}\leq \frac{\hbar^2}{2m}\frac{1}{c(g)A(\Sigma)}\cdot$$ 
 We have to remark here that the energy levels are continuous (non-discrete energy levels) in the classical limit  of $\hbar\to 0$ or in the high mass limit of $m\to \infty$, and for any given compact surface $\Sigma\subset \mathbb{R}^3$ if we consider the re-scaled family of surfaces $\{\lambda \Sigma \}_{\lambda>0}$, then $$\lim_{\lambda\to\infty}E_k(\lambda \Sigma)-E_0(\lambda\Sigma)=0.$$ In short, the classical limit  is obtained at large mass or large scale -- as expected.
 
\section{Surfaces of  constant skew curvature}
 The induced Schrödinger \eqref{extrinsic} equation leads to the study of surfaces of constant skew curvature. From a physical point of view, a constant wave function, $\Psi=c$, is a stationary solution of the Schrödinger equation if there is a value of $E$ such that $-\frac{1}{4}S^2c = \frac{2mE}{\hbar^2}c$, which means that the surface has constant skew curvature.  For a compact surface, if the particle is initially in a uniformly distributed wave function, $\Vert \Psi\Vert^2=\frac{1}{{\rm A}(\Sigma)}$, it will remain so if it is an eigenfunction of the Hamiltonian:
$$
\Vert \nabla \Theta\Vert^2+i\Delta\Theta=\frac{1}{4}S^2+\frac{2m E}{\hbar^2}.
$$
The necessary condition for this to occur is that the surface has constant skew curvature. Because in a uniformly distributed wave function, we have a real function $\Theta: \Sigma\to\mathbb{R}$ such that 
$$\Psi=\frac{e^{i \Theta}}{\sqrt{{\rm A}(\Sigma)}}\cdot$$
Then
$$
\Vert \nabla \Theta\Vert^2+i\Delta\Theta=\frac{1}{4}S^2+\frac{2m E}{\hbar^2},
$$
which implies that $\Theta$ is a harmonic function ($\Delta \Theta=0$). But the only harmonic functions in a compact surface is the constant function, and thus the surface must have constant skew curvature. On the other hand, if the compact surface has constant skew curvature, the geometrically induced potential in this case is constant, leading to the spectrum of energies coinciding with the pure Laplace spectrum, with the ground state being the uniformly distributed wave function itself. To obtain a non-uniformly distributed wave function, a certain amount of energy must be supplied, the size of which is determined by our results. Indeed, from \eqref{result1} and assuming constant skew curvature $S=c$ on $\Sigma$ and $\Sigma$ to be an oriented  compact surface of genus $g$, and by using the Gauss-Bonnet theorem we have
\begin{equation}\label{eq:oka}
\frac{\hbar^2}{m}\frac{\mathcal{W}(\Sigma)}{{\rm A}(\Sigma)}=     \frac{\hbar^2}{m}\left(c^2+\frac{4\pi(1-g)}{A(\Sigma)}\right)\geq    E_1-E_0.
\end{equation}
The bounds in the gap spectrum of energy are improved when the surface has constant skew curvature. The gap in the spectrum of energy is given by the difference between the first two energy levels and can be bounded by the ratio of the Willmore functional to the area of the surface. In this specific case, with constant geometrically induced potential, since the energy spectrum can be deduced from the Laplace spectrum, by using Theorem 1 of \cite{MR577325}, we can provide the following improvement of inequality \eqref{eq:oka}:
\begin{equation}\label{eq:nona}
E_1-E_0     \leq \frac{\hbar^2}{m A(\Sigma)}\cdot \min\left\lbrace4\pi(1+g), \mathcal{W}(\Sigma)\right\rbrace.
\end{equation}
In our main results for upper bounds in the gaps in the spectrum of energy, inequalities \eqref{result1} and \eqref{result2}, we always have the term  
$$
\frac{1}{4}\left(\max_{\Sigma}S^2-\overline{S^2}\right),
$$
which vanishes exactly when the surface has constant skew curvature. What we have stated then shows that 
 surfaces of constant skew curvature play an important role in understanding the behavior of a particle confined to a surface. They are the only surfaces where stationary solutions with uniformly distributed wave function are allowed and moreover, they are the unique surfaces where equality could be attained for our upper bounds for the first gap in the energy spectrum.
 
Surfaces of constant skew curvature have attracted much attention, see \cite{daSilva2017}, \cite{Toda201751} and \cite{Lopez2020} for instance. In fact,  R. López and A. Pámpano have recently classified the surfaces of revolution with constant skew curvature, see \cite{Lopez2020}.
One important result from this classification is that there are no other regular and compact surfaces of revolution with constant skew curvature than the round spheres. \\

This motivates the following pertinent open problem:

\begin{quote}
    Are the round spheres the only compact regular surfaces without boundary and with constant skew curvature that are immersed in Euclidean $3$-space?
\end{quote}

 % 

%%%%%%%%%%%%%%%%%%%%%%%%%%%%%%%%%%%%%%%%%%%%%

This geometric problem is similar to the classical geometric problems of constant curvature. For example, in 1853, J.H. Jellet demonstrated that if a compact, star-shaped surface in 3-dimensional space, $\mathbb{R} ^{3}$, has a constant mean curvature, it is isometric to a standard round sphere.  Alexandrov later confirmed that any compact, embedded surface in the Euclidean $3$-dimensional space with a non-zero constant mean curvature must also be a sphere. The assumption on the embeddedness cannot be removed. Indeed, in 1984, Henry C. Wente created the Wente torus, which is an immersion of a torus with constant mean curvature in $\mathbb{R} ^{3}$. In the same vein it is well known that a compact simply connected surface with constant Gaussian curvature must also be a round sphere.

%%%%%%%%%%%%%%%%%%%%%%%%%%%%%%%%%%%%%%%%%%%%%%
Surfaces of revolution that have a constant, non-zero skew curvature constitute a family of surfaces parametrized by a single parameter. As already alluded to above, there are no compact regular surfaces of revolution with constant non-zero skew curvature. However, in the non compact case regular surfaces of revolution do appear, see \cite{Lopez2020}.

For a brief illustration of this point, consider a surface of revolution with the following parametrization for $t\geq 0$ and generator function $g(t)$:

\begin{equation} \label{eq:r1}
r(t,u) = (t \cdot \cos(u), t \cdot \sin(u), g(t))\quad .
\end{equation}

The absolute value of the difference between the two principal curvatures, i.e. the skew curvature of the surface, is then:
\begin{equation} \label{eq:S1}
S(t) = \frac{-h^{3}(t) + t\cdot h'(t) - h(t)}{2 \left( 1+ h^{2}(t) \right)^{3/2}} \quad , 
\end{equation}
where $h(t) = g'(t)$.
The solution $h(t)=0$ to $S(t) = 0$ gives horizontal planes. The non-zero solutions to
$-h^{3}(t) + t\cdot h'(t) - h(t) = 0$ are the spheres:
\begin{equation}
h(t) = \pm\frac{t}{\sqrt{\alpha - t^{2}}} \quad,
\end{equation}
where $\alpha$ is an integration constant.

\begin{figure}[h!]
\includegraphics[height=50mm]{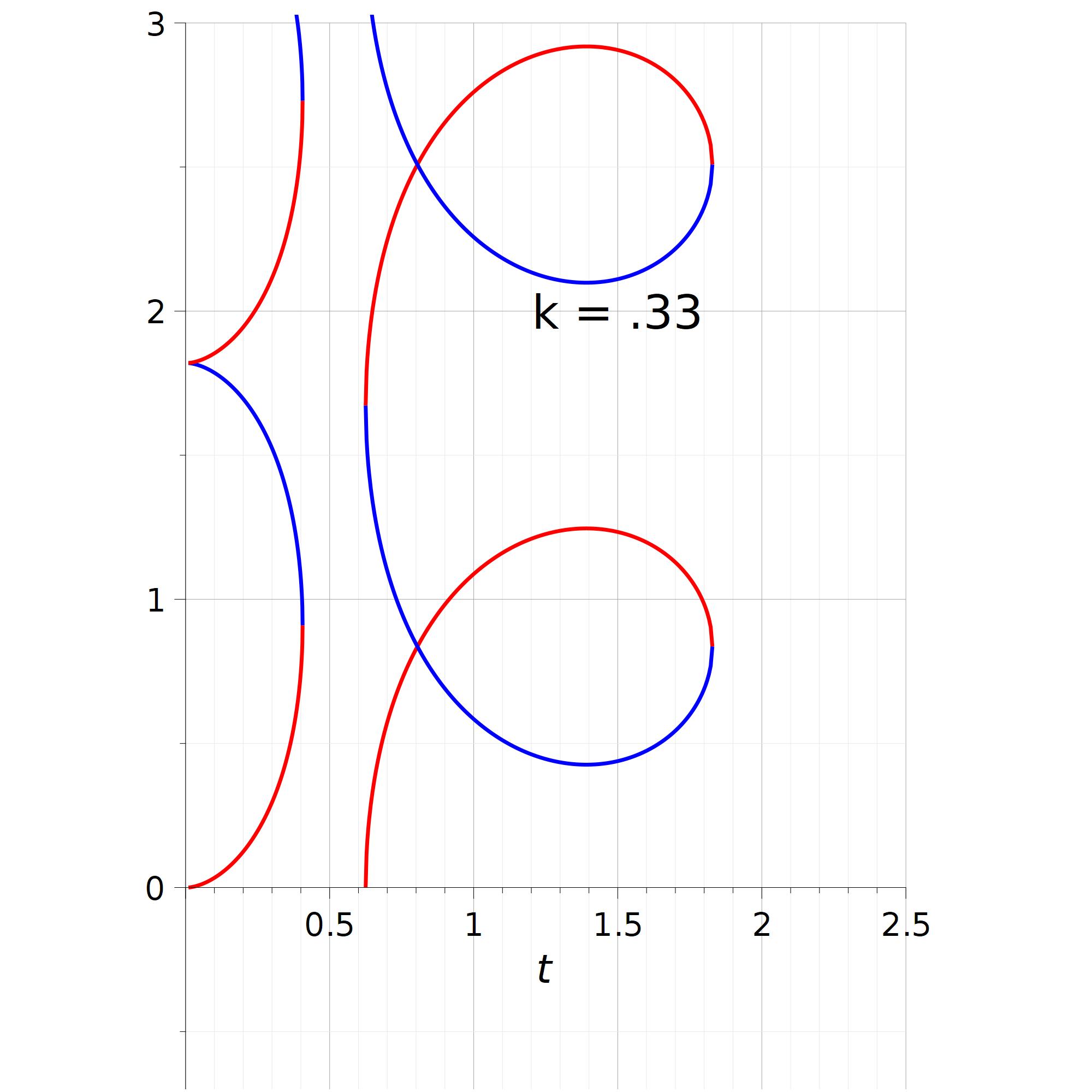} \quad  \includegraphics[height=50mm]{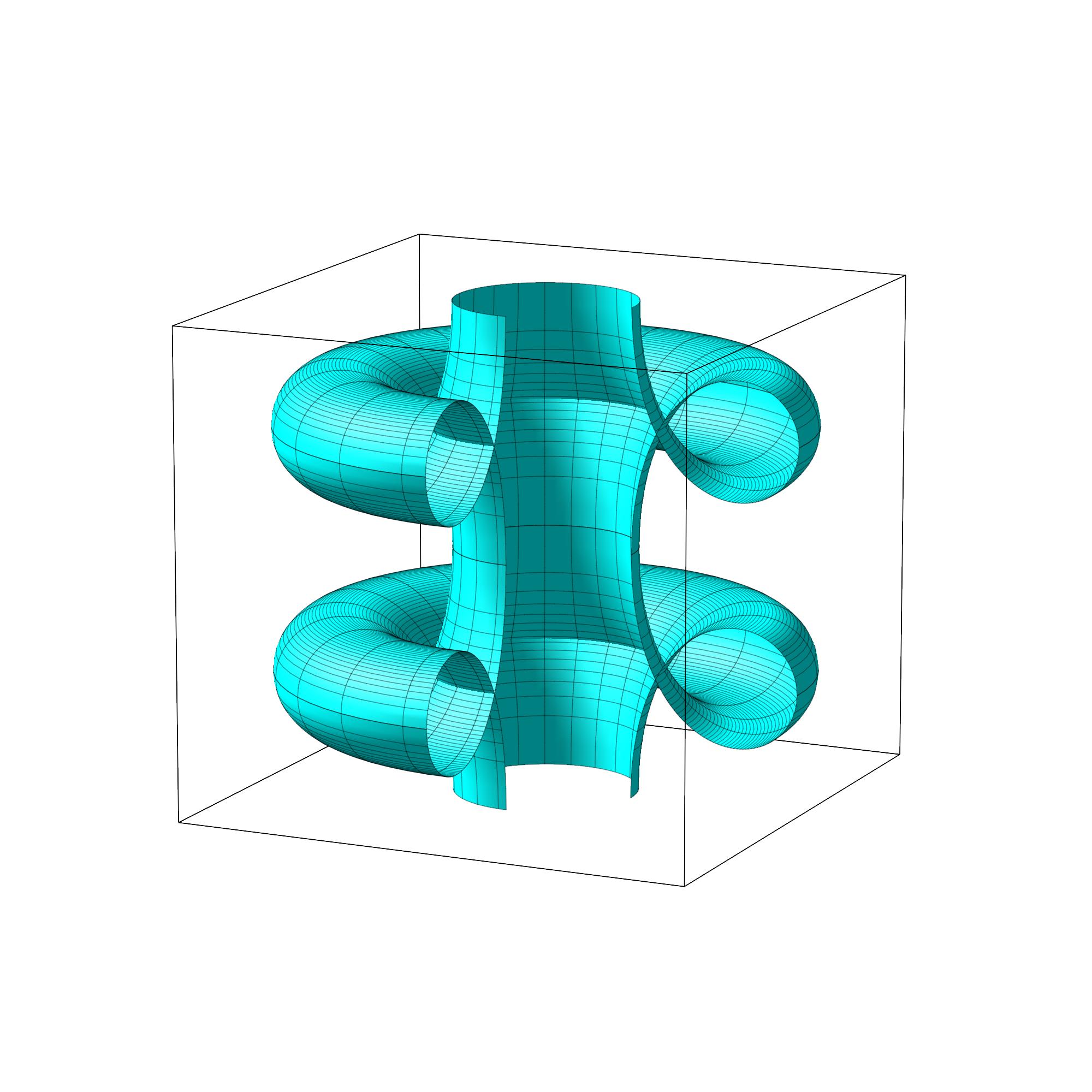}
\caption{\small{
Profile curves for surfaces of revolution with constant skew curvature $1$ and $k = 0.33 >  k_{0}$.
The curves are obtained by periodically stacking the (red) base curve solutions (with $h(t) \geq 0$) and their symmetric (blue) companions (with $h(t) \leq 0$). Only the outer-most stacking is smooth. The corresponding outer-most surface is also shown with two periods only. That surface is compact in $\mathbb{R}^{3}/\mathbb{Z}$ as discussed below.}}\label{fig:1}
\end{figure}

\begin{figure}
\includegraphics[height=50mm]{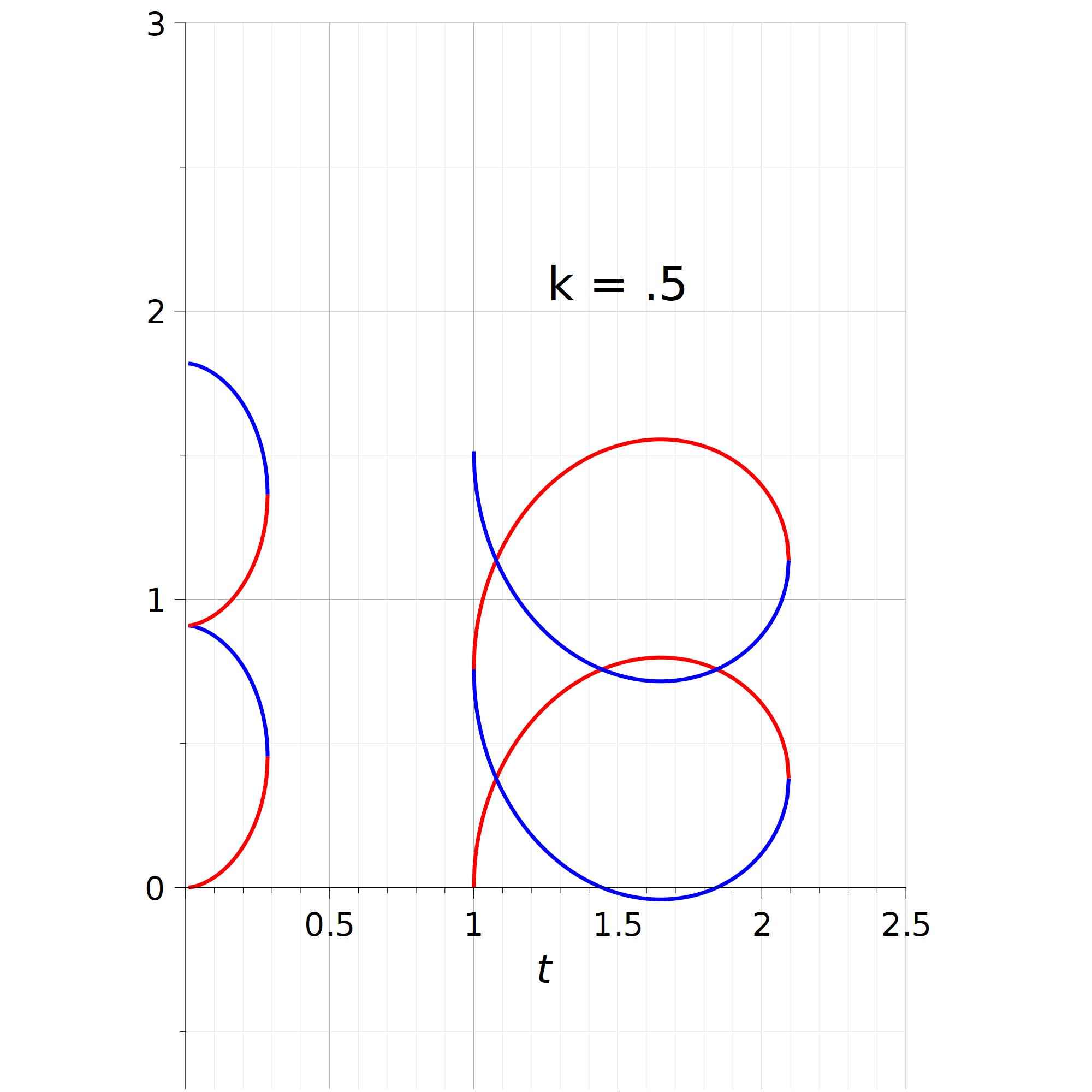} \quad  \includegraphics[height=35mm]{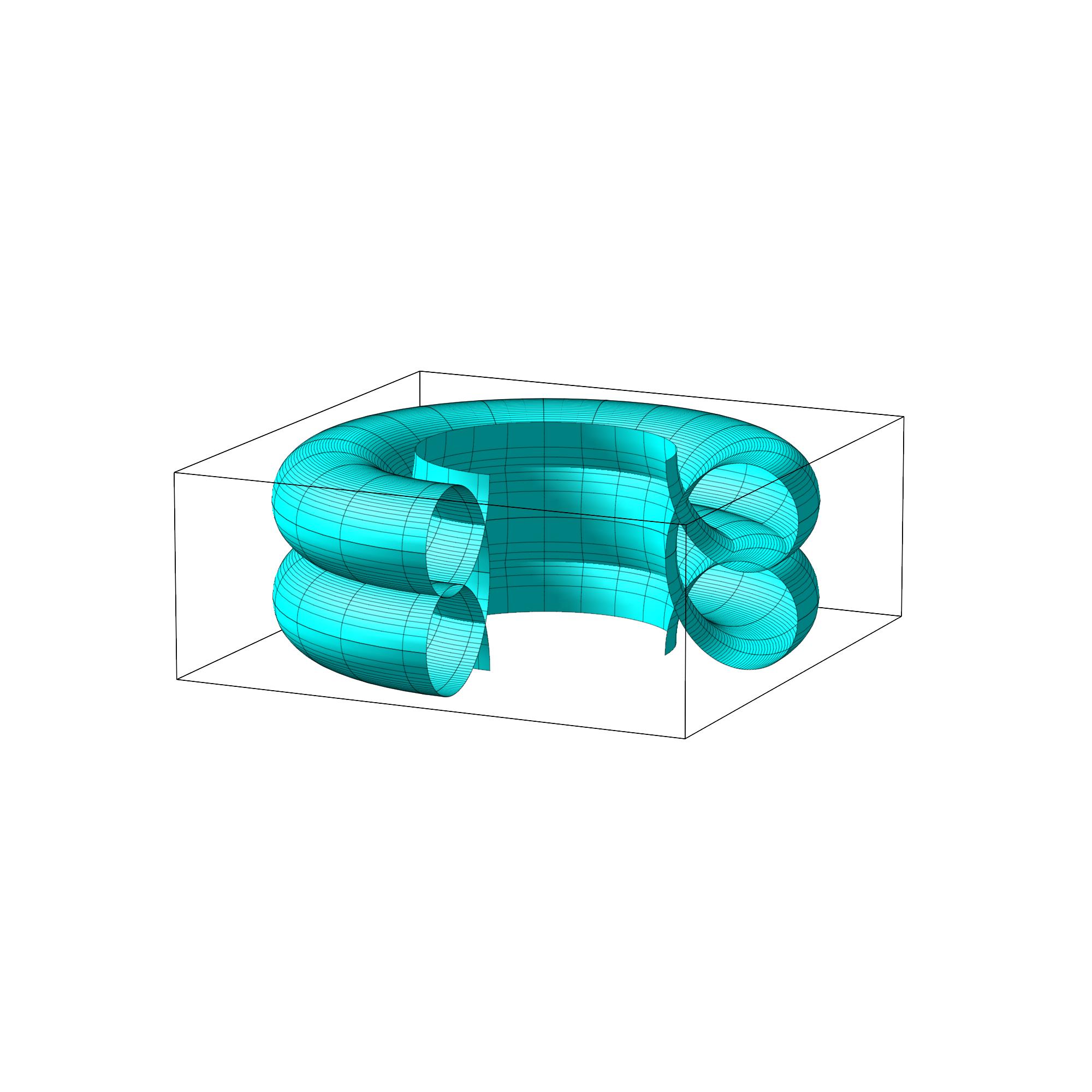}
\caption{\small{Similar to figure \ref{fig:1}, except here with $k=0.5> k_{0}$ .}}\label{fig:2}
\end{figure}

\begin{figure}
\includegraphics[height=50mm]{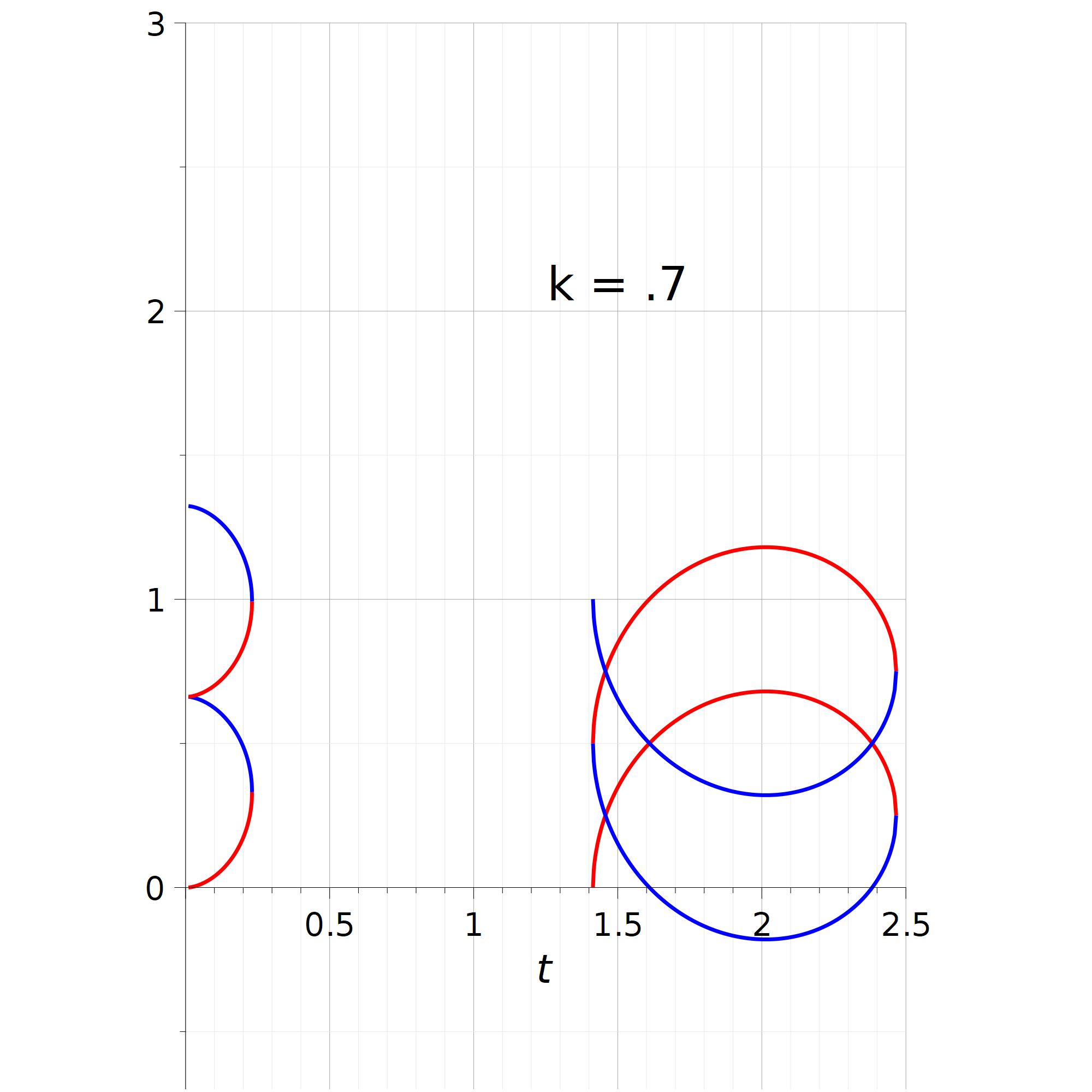} \quad  \includegraphics[height=30mm]{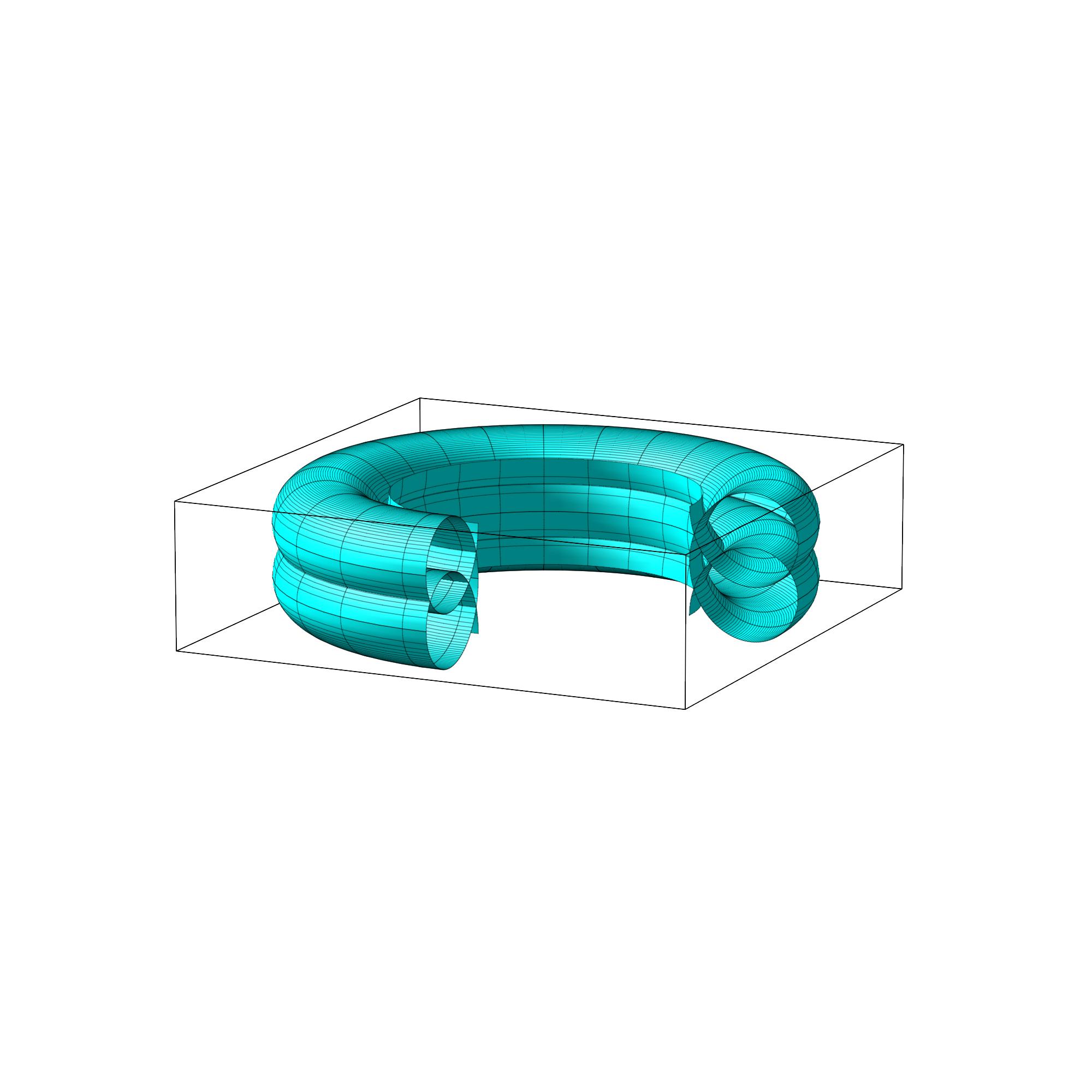}
\caption{\small{Similar to figure \ref{fig:1}, except here with $k=0.7> k_{0}$ .}}\label{fig:3}
\end{figure}

For constant non-zero skew curvature we only need to consider $S(t)=1$ since all other cases appear by scaling. 
Such surfaces (choosing $h(t) \geq 0$, which corresponds to the red curves in the Figures below) satisfy
\begin{equation}
\ln(t) + \frac{h(t)}{t \cdot \sqrt{1 + h^{2}(t)}} = k \quad,
\end{equation}
where $k$ is a constant of integration. We solve for $h(t)$:
\begin{equation} \label{eq:h(t)}
h(t) = \frac{2 ( k -\ln(t)) \cdot t}{\sqrt{1+ (8 k -4 \ln(t))\cdot t^{2} \cdot \ln(t) -4 k^{2}\cdot t^{2}}} \quad .
\end{equation}
And thence, for each allowed $k$ we obtain $g(t)$ by (numerical) integration of $h(t)$. For each value of $k$ there are positive $t$-intervals for which the square root in the denominator is well defined in \eqref{eq:h(t)}. There are at most 3 roots of the denominator, i.e. there are at most two $t$-intervals to consider. There is one such interval (i.e one constant skew curvature surface) for $k<k_{0}$ and two intervals (i.e. two constant skew curvature surfaces) for $k>k_{0}$ where $k_{0}$ is the bifurcation value corresponding to the vertical cylinder of radius $1/2$:
\begin{equation}
k_{0} = 1 - \ln(2) \approx 0.307 \quad .
\end{equation}
We integrate (numerically) the functions $h(t)$ for each value of $k$ and obtain the corresponding generator functions $g(t)$ and thus the corresponding surfaces of revolution. See figures \ref{fig:1}, \ref{fig:2}, \ref{fig:3}, where the results are shown for three values of $k > k_{0}$. The corresponding 'outer' solutions -- which do not intersect the axis of revolution -- are the most interesting here, since they can be stacked periodically to give smooth non-compact surfaces with constant skew curvature.
%\begin{figure}[h!]
%\includegraphics[height=40mm]{FuncFigNew97} %\includegraphics[height=40mm]{FuncFigNew98}
%\includegraphics[height=40mm]{FuncFigNew99} \\
%\includegraphics[height=40mm]{FuncFigNew100} %\includegraphics[height=40mm]{FuncFigNew101}
%\includegraphics[height=40mm]{FuncFigNew102}
%\caption{\small{XX}}
%\end{figure}


\section{Upper bounds for the spectral gap via geometric compactification}

The techniques of this paper aim to derive upper bounds for the first gap in the spectrum for compact surfaces immersed in the Euclidean space. However, for non-compact surfaces with certain symmetries, some -- or all -- of our results can still be adapted.

One specific scenario where our technique can be applied is for non-compact surfaces of revolution with a profile curve that has both a translational symmetry and a reflection symmetry. This is precisely the case for the profile curves shown in figures \ref{fig:1}, \ref{fig:2}, and \ref{fig:3} and thence the associated surfaces of revolution.

These surfaces are actually immersed surfaces in $\mathbb{R}^3$, obtained as the image of an immersion $\varphi: M=\mathbb{R} \times \mathbb{S}^1 \to \mathbb{R}^3$. The surface is said to have a translation symmetry if there exists a vertical vector $\vec{A} \in \mathbb{R}^3$ such that the surface is invariant under translation by $\vec{A}$. Additionally, it is said to have a reflection symmetry around the $z$ axis if the surface is invariant under reflection about this axis.

We say that $\Sigma=\varphi(M)$ admits a translation symmetry because there is a vertical vector $\vec{A}\in \mathbb{R}^3$ such that the surface sigma is invariant under the translation of $\vec{A}=(0,0,A)$,
$$
T_{\vec{A}}(\Sigma):=\{x+\vec{A} \, :\, x\in \Sigma\}=\Sigma.
$$
Moreover, we say that $\Sigma$ admits a  reflection $R_z$ around the $z$ axis because,
$$
R_z(\Sigma):=\{(x_1,x_2,-x_3):\, (x_1,x_2,x_3)\in \Sigma\}=\Sigma.
$$

When the profile curve is arc-length parametrized, $M$ can be equipped with global coordinates $s$ and $\theta$, where $s$ is the arc-length parameter and $\theta$ is the corresponding rotation angle. The Hamiltonian operator $\widetilde{H}=\frac{\hbar^2}{2m}\left[-\Delta -\frac{1}{4}\left(k_1-k_2\right)^2\right]$ is symmetric with respect to a translation $\tau_a (s,\theta)=(s+a,\theta)$ and a reflection parity change $P(s,\theta)=(-s,\theta)$. Therefore, there exists  unitary operators $\widetilde{\tau}_a$ and  $\widetilde{P}$ such that 
$$
[\widetilde{H},\widetilde{\tau}_a]=0, \quad [\widetilde{H},\widetilde{P}]=0.
$$
However, it is not possible to obtain a basis of simultaneous eigenstates for $\widetilde{H}$, $\widetilde{\tau}_a$, and $\widetilde{P}$, because $\widetilde{\tau}_a$ and $\widetilde{P}$ do not commute.

We are interested in the composition $T_a = \tau_a \circ P$, which has the property that $T_a \circ T_a = I$. This allows us to obtain a simultaneous basis of eigenstates for $\widetilde H$ and $\widetilde T_a$, with eigenvalues of $\widetilde T_a$ being $\pm 1$. Therefore, any eigenfunction of $\widetilde H$ can be decomposed into eigenfunctions of $\widetilde T_a$ with eigenvalues $\pm 1$.
$$
\Psi_E(s,\theta)=A \Psi^{+}_E(s,\theta)+B \Psi^{-}_E(s,\theta),\quad A,B\in \mathbb{C},
$$
where $$\widetilde T_a^2\Psi^{+}_E(s,\theta)=\Psi^{+}_E(s,\theta),\quad \widetilde T_a^2\Psi^{-}_E(s,\theta)=-\Psi^{-}_E(s,\theta).$$
Then, when we look for the wave functions
$$
\langle s,\theta\vert \widetilde T_a \vert E_i, \epsilon_i\rangle=\epsilon_i \Psi_{E_i,\epsilon_i}(s,\theta), 
$$
But since $\widetilde T_a\vert s,\theta\rangle=\vert -s+a,\theta\rangle$ we can conclude that
\begin{equation}
\Psi_{E_i,\epsilon_i}(-s+a,\theta)=\epsilon_i \Psi_{E_i,\epsilon_i}(s,\theta). 
\end{equation}
Taking $s=0$ we obtain
$$
\Psi_{E_i,\epsilon_i}(a,\theta)=\epsilon_i \Psi_{E_i,\epsilon_i}(0,\theta).$$ 
Hence, in the case $\epsilon_i=1$,
\begin{equation}\label{periodic}
\Psi_{E_i,1}(a,\theta)=\Psi_{E_i,1}(0,\theta).    
\end{equation}

In summary, the spectrum $\sigma(\Sigma)$ of $\Sigma$ can be divided into two parts: one, denoted as $\sigma^{+}(\Sigma)$, is related to wave functions of positive parity with respect to $\widetilde{T}_a$, and the other, denoted as $\sigma^{-}(\Sigma)$, is related to wave functions of negative parity with respect to $\widetilde{T}_a$. If $\Sigma$ has positive parity, the wave function must satisfy the boundary value problem \eqref{periodic}.

From a geometric perspective, the spaces $\mathbb{R}^3$ and $M$ can be divided into equivalence classes using the relation of equivalence $(x,y,z)\sim (x',y',z')$ in $\mathbb{R}^3$  if and only if there exist $k\in \mathbb{Z}$ such that 
$$
(x,y,z)-(x',y',z')=k\vec{A},
$$
and $(s,\theta)\sim (s',\theta')$ in $M$ if and only if, there exist $k\in \mathbb{Z}$ such that 
$$
T^k_a(s,\theta)=(s',\theta').
$$ 
The quotient spaces will be denoted as $\mathbb{R}^3/\mathbb{Z}$ and  $M/\mathbb{Z}$, respectively.   Since the translations act by isometries we can use the following immersion $\beta:M/\mathbb{Z}\to \mathbb{R}^3/\mathbb{Z}$ by using the following commutative diagram:
\[ \begin{tikzcd}
M \arrow{r}{\varphi} \arrow[swap]{d}{\pi} & \mathbb{R}^3 \arrow{d}{\pi} \\%
M/\mathbb{Z}\arrow{r}{\beta}& \mathbb{R}^3/\mathbb{Z}
\end{tikzcd}
\]
where $\pi$ denotes the canonical map to the classes of equivalence, $x\mapsto \pi(x):=[x]$. Because the spaces $M$ and $M/\mathbb{Z}$ and the spaces $\mathbb{R}^3$ and $\mathbb{R}^3/\mathbb{Z}$ are locally isometric, since $\varphi$ is an isometric immersion, $\beta$ is an isometric immersion as well with the same values of $H$ and $K$.

The immersion $\beta:M/\mathbb{Z}\to \mathbb{R}^3/\mathbb{Z}$ can be used to obtain the first gap in the spectrum of $\Sigma$ with positive parity. If $M/\mathbb{Z}$ is smooth enough, this gap is bounded by the following:
$$
E_1-E_0\leq \frac{\hbar^2}{2m}\left(\frac{2\mathcal{W}(\beta(M/\mathbb{Z}))}{{\rm A} (\beta(M/\mathbb{Z}))}+\frac{1}{4}\left(\max_{\beta(M/\mathbb{Z})} S^2-\overline{S^2}\right)\right)
$$
for $E_1$ and $E_0$ in $\sigma^{+}(\Sigma)$. If the surface $\Sigma$ is constructed as the translation of a fundamental domain $L$ and has a reflection symmetry, the first gap in the spectrum of positive parity is bounded by:
$$
E_1-E_0\leq \frac{\hbar^2}{2m}\left(\frac{2\mathcal{W}(L)}{{\rm A} (L)}+\frac{1}{4}\left(\max_{L} S^2-\overline{S^2}\right)\right).
$$
In the case of constant skew curvature, this becomes:
$$
E_1-E_0\leq \frac{\hbar^2}{2m}\frac{2\mathcal{W}(L)}{{\rm A} (L)}.
$$
Observe that this upper bounds proportional to the square mean curvature are similar to the bounds obtained in \cite{Gravesen2010}, where truncated surfaces of revolution are used with general boundary
conditions (Dirichlet, Neumann, or Robin).

In the case shown by figures \ref{fig:1}, \ref{fig:2} and \ref{fig:3}, since $M/\mathbb{Z}$ is a torus  with constant skew curvature $S=1$, by using the Gauss-Bonnet theorem,
$$
{\rm A}(M/\mathbb{Z})=\int_{M/\mathbb{Z}}S^2dA=\int_{M/\mathbb{Z}}H^2dA=\mathcal{W}(M/\mathbb{Z}),
$$
and therefore
$$
E_1-E_0\leq \frac{\hbar^2}{m}.
$$
\section{Conclusions}
The main objective of this study was to investigate which geometric or topological properties play a significant role in describing a particle confined to a compact surface. With a compact surface possessing a bounded effective and geometrically induced potential, the energy levels are discrete. Determining the energy spectrum of a particle confined to a surface can be a challenging task, even numerically. However, this study demonstrates that the difference between the  energy of the first excited state and the ground level  can be upper-bounded by two terms. One of these terms is related to the total area and the Willmore energy, or some topological constant that depends only on the genus of the surface. The other term measures how far the surface deviates from being a surface of constant skew curvature.

This leads naturally to a study of the properties of surfaces with constant skew curvature, specifically the surfaces of revolution of constant skew curvature. They constitute a one-parameter family with a bifurcation point between the compact and non-compact cases. Although the results in this report are stated for the compact case, if a non-compact surface possesses enough symmetries, the results can still be used via the shown compactification of the surface to obtain bounds for a specific part of the original surface's spectrum.

\section*{Acknowledgments}This Work was partially supported by the Research grant  PID2020-115930GA-100 funded by MCIN/ AEI /10.13039/501100011033,  and AICO/2021/252. 
The authors would like to thank Luiz C. B. da Silva for comments and suggestions that have improved the presentation of the present work.
The authors would also like to acknowledge the hospitality of the departments of mathematics at DTU and at Universitat Jaume I, respectively.

\end{document}